\documentclass[sigconf, nonacm]{IEEEtran}

\usepackage{cite}
\usepackage{amsmath,amssymb,amsfonts}
\usepackage{algorithmic}
\usepackage{graphicx}
\usepackage{textcomp}
\usepackage{xcolor}
\usepackage[colorlinks,linkcolor=blue]{hyperref}
\usepackage[justification=centering]{caption}
\usepackage{setspace}
\usepackage{enumerate}
\usepackage{float}
\usepackage{CJKutf8}

\begin{document}
\title{Stock Market Trend Analysis Using Hidden Markov
Model and Long Short Term Memory}


\author{
\IEEEauthorblockN{
Mingwen Liu$^{a}$ $^{*}$\thanks{*Corresponding author. Email Address: liumwen@mail2.sysu.edu.cn},
Junbang Huo$^{b}$, 
Yulin Wu$^{b}$, 
Jinge Wu$^{c}$\\
}

\IEEEauthorblockA {
  $^{a}$ Business School, Sun Yat-sen University, Guangzhou, PR China\\
}

\IEEEauthorblockA {
  $^{b}$ School of Mathematics, Sun Yat-sen University, Guangzhou, PR China\\
  }
  
  \IEEEauthorblockA {
  $^{c}$ School of Science, Xi'an Jiaotong-Liverpool University, Suzhou, PR China\\
}
}

\maketitle
\thispagestyle{empty}

\begin{abstract}
This paper intends to apply the Hidden Markov
Model into stock market and and make predictions. Moreover, four different methods of improvement,
which are GMM-HMM, XGB-HMM, GMM-HMM+LSTM and
XGB-HMM+LSTM, will be discussed later with the results of
experiment respectively. After that we will analyze the pros and
cons of different models. And finally, one of the best will be used
into stock market for timing strategy.
\end{abstract}



\section{Introduction}

HMM and LSTM have been widely used in the field of
speech recognition in recent years, which has greatly improved
the accuracy of existing speech recognition systems. Based
on many similarities between speech recognition and stock
forecasting, this paper proposed the idea of applying HMM and LSTM
into financial markets, and used machine learning algorithms to
predict ups and downs of stock market. We explored the effect
of the model through experiments. First, we use the GMM-HMM
hybrid model method. Then, improve HMM model by
using XGB algorithm (XGB-HMM). Next, we build a longshort
term memory model (LSTM), then use GMM-HMM and
LSTM together, also compared results with XGB-HMM and
LSTM. By comparing the results of the four experiments, the
model which is most suitable for the stock market wil be
summarized and applied. This paper will explain the core of
each algorithm and the practical operation, and the core codes of experiments will be open source for reference and learning on GitHub.\footnote{\href{https://github.com/JINGEWU/Stock-Market-Trend-Analysis-Using-HMM-LSTM}{https://github.com/JINGEWU/Stock-Market-Trend-Analysis-Using-HMM-LSTM}}.

\section{Data Process}

\subsection{\underline{Feature}}
\subsubsection{Construction of Y Feature}
\begin{figure}[ht]
	\centering
		 
	\includegraphics[scale=0.5]{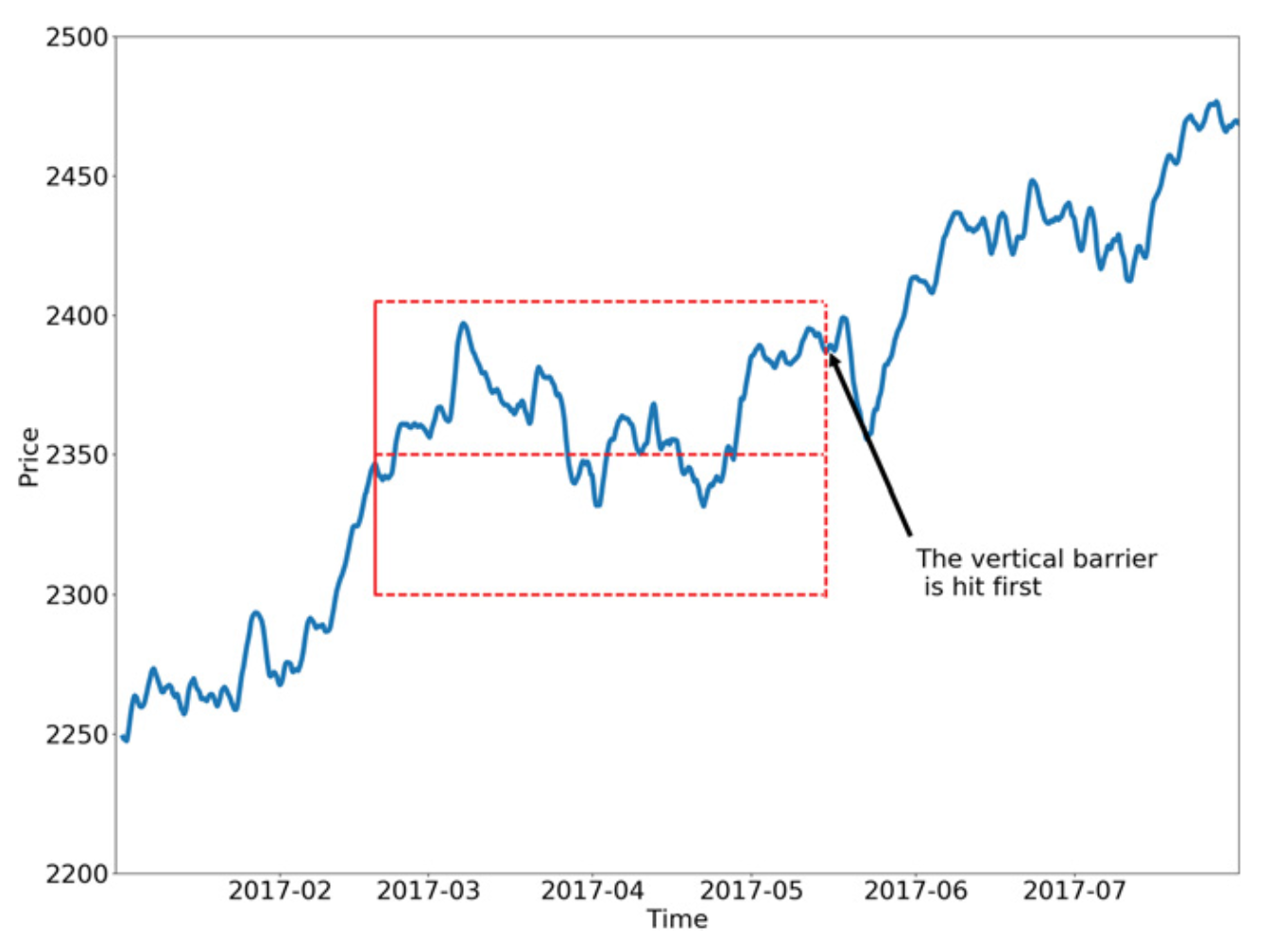}
    \caption{the triple barrier method [\hyperref [ref 5]{5}] }
	\label{fig 1}
\end{figure}

First, we need to create a feature Y which can reflect the trend of stock market in order to
predict the ups and downs of stock price. Thus a method
named the triple barrier method[\hyperref[ref 5]{5}] can be well used in this situation.The triple barrier method is put forward by Marcos
for the first time, which is an alternative labelling method (see
in Fig. \ref{fig 1}).

\vspace{0.5cm}

\begin{table}[!ht]
	\centering
	\begin{tabular}{ll}
		\hline
		\textbf{Notation}\\
		\hline
		T : length of the observation sequence\\
		N : number of states in the model \\
		M : number of observation symbols\\
		A : state transition probabilities\\
		B : emission probability matrix \\
		$ \pi $ : initial state distribution \\
		$ O =\{O_0,O_1,..,O_{T-1}) \}$ : observation sequence \\ 
		$ S =\{ S_0,S_1,...,S_T{T-1} \}$ : state sequence \\ 
		$ V =\{ V_0,V_1,...,V_{M-1}\} $ : set of possible observations \\
		$\lambda =(A,B,\pi) $ : HMM model \\
		$socre\_plot $ : score function \\
		\hline\\
	\end{tabular}
	\caption{Notation}
	\label{Table 1}
\end{table}
\vspace{0.5cm}

\begin{figure}[ht]
	\centering
	\includegraphics[scale=0.28]{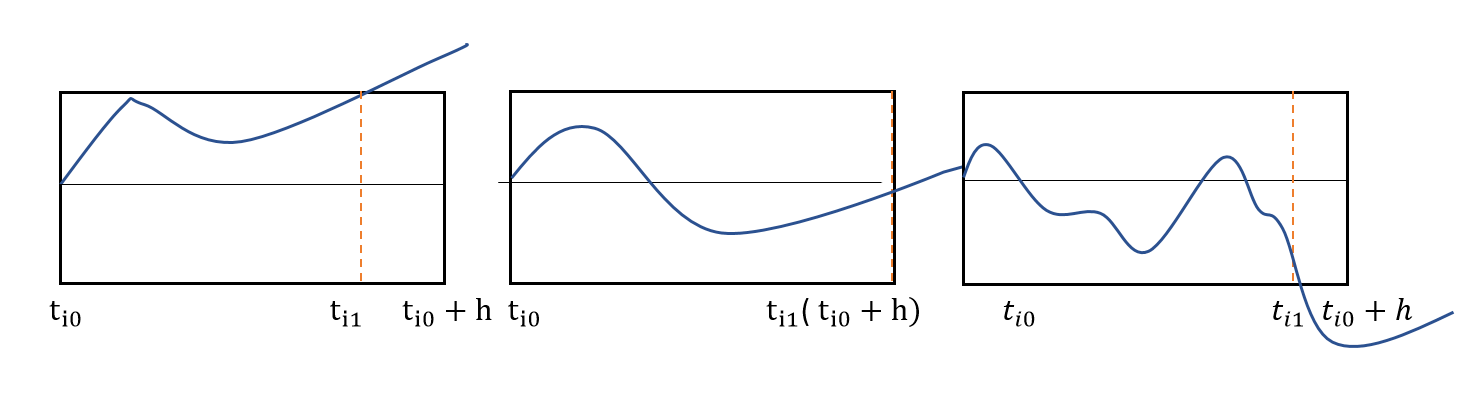}
	\caption{three different status of the triple barrier method [\hyperref [ref 5]{5}] } \label{fig 2}
\end{figure}

It sets three dividing lines and marks them according to
the first barrier touched by the path. First, we set two horizontal barriers and one vertical barrier. The two horizontal barriers are defined by profit-taking and stop loss limits, which are a dynamic function of estimated volatility (whether realized or implied). The third barrier is defined in terms of number of bars elapsed since the
position was taken (an expiration limit). If the upper barrier
is touched first, we label the observation as a 1. If the lower
barrier is touched first, we label the observation as a -1. If the
vertical barrier is touched first, we label the
observation as 0 (shown in Fig. \ref{fig 2}). The triple barrier method can be helpful to label stock's future price.

One problem with the triple-barrier method is path-dependent. In order to label an observation, we need to take the entire path spanning $ [ t_{i0},t_{i0} + h ] $ into account, where $h$ defines the vertical barrier (the expiration limit). We denote $t_{i1}$  as the time of the first barrier touch, and the return associated with the observed feature is $t_{i0},t_{i1}$. For the sake of clarity, $t_{i1} \leq t_{i0} + h$, and the horizontal barriers are not necessarily symmetric.
\subsubsection{Construction of the Observation Sequence}
Our data of observation sequence is divided into 8 different types:

\begin{table}[!ht]
	\centering
	\scalebox{0.8}{
	\begin{tabular}{ll}
		\hline
Type & Meaning \\
\hline
Market Factor & pre-close price,open price
close price,\\
&deal amount etc.\\

\hline
Quality factor & 50 factors such as accounts payable ratio,\\
&turnover days, management expenses\\
&and total operating income\\
\hline
Income risk Factor & 10 factors such as Cumulative Income and\\ 
&falling fluctuation\\

\hline
Value Factor & 13 factors such as
cash flow market value ratio,\\
&income market value ratio, price-to-book ratio\\
&and price-earnings ratio\\

\hline
Mood Factor & 49 factors such as hand turnover rate\\
&dynamic trading and trading volume\\
\hline
Index Factor& 49 factors such as exponential moving average\\
&and smoothing similarity moving average\\

\hline
 Momentum Factor & 56 factors such as stock income\\
&and accumulation/distribution line\\

\hline
Rise Factor & 17 factors such as net asset growth rate\\
&and net profit growth rate\\

		\hline\\
	\end{tabular}
	}
	\caption{Features}\label{Table 2}
\end{table}

In order to classify different types of observation sequence, we use a score function to evaluate these features and choose powerful ones into model. Please refer to TABLE \ref{Table 1} as the declaration of notations used in this paper.

Fix one factor F, assume the daily data of F is $O$, train GMM-HMM model, and generate the optimal state sequence $S$ by using Viterbi algorithm.

Create a count matrix M, where M$ \in R^{N \times 3} $,in which $M_{ij}$ represent the frequency of label $j$ corresponding to state $i$, where $ i \in \{0,1,...,N-1\}, j\in \{-1,0,1\} $
Constructing a counting ratio matrix $MR$ according to M where,
$$MR_{ij} = \frac{M_{ij}}{\sum_{j}M_{ij}}$$
The accuracy of state $i$ is
\begin{equation}
Acc_{i} = max_{j} \{MR_{ij}\}
\end{equation}
where $$i \in \{0,1,...,N-1\},i \in \{0,1,...,N-1\},j\in \{-1,0,1\} $$
The entropy of state $i$ is
\begin{equation}
H_i = -\sum_{j} MR_{ij} log(MR_{ij}) 
\end{equation}
where $$i \in \{0,1,...,N-1\},i \in \{0,1,...,N-1\},j\in \{-1,0,1\} $$
The weights of state $i$ is
\begin{equation}
w_i  = \frac{\sum_{i} M_{ij}  }{\sum_{i}\sum_{j} M_{ij}}
\end{equation}
And therefore the score function can be described as
\begin{equation}
\text{score} = \text{score}(Acc,H,w) = \sum_{i} (Acc_i  \times \frac{1}{1+H_i} \times w_i) 
\end{equation}
We consider the higher the score is, the better the feature is.

After computing, the features selected by score function is in TABLE \ref{Table 3}.

\begin{table}[!ht]
\begin{spacing}{1.1}

	\centering
	\scalebox{0.7}{
	\begin{tabular}{ll}
		\hline
Type & Meaning \\
\hline
Market Factor & $\text{log\{ Today's Close Price} -(T-5) \text{ day's Close Price}\}$\\
&$\text{log}(\frac{\text{Highest Price}}{\text{Lowest Price}})$\\
&$\text{log\{ Today's Close Price} -(T-5) \text{ day's Close Price}\}$\\
&$\frac{\text{Close Price}}{\text{Pre Close Price}}$\\
&$\frac{\text{Open Price}}{\text{Pre Close Price}}$\\
&$\frac{\text{Open Price}}{\text{Pre Close Price}}$\\
&$\frac{\text{Highest Price}}{\text{Pre Close Price}}$\\
&$\frac{\text{Lowest Price}}{\text{Pre Close Price}}$\\

\hline
Quality factor & DEGM, EBITToTOR, MLEV, CFO2EV,\\
&NetProfitRatio , ROA5\\
\hline
Income risk Factor & DDNBT, TOBT\\

\hline
Value Factor & CTP5, SFY12P\\

\hline
Mood Factor & ACD20, ACD6, STM, VOL5, DAVOL5\\
\hline
Index Factor& MTM, UpRVI, BBI, DownRVI, ASI\\

\hline
 Momentum Factor & BBIC, PLRC6, ChandeSD, PLRC12, ChandeSU,\\ &BearPower\\

\hline
Rise Factor & FSALESG , FEARNG\\

		\hline\\
	\end{tabular}
	}
	\caption{Filtered features}
	\label{Table 3}
	
\end{spacing}
\end{table}

\begin{figure*}[ht]
	\centering
	\includegraphics[scale=0.33]{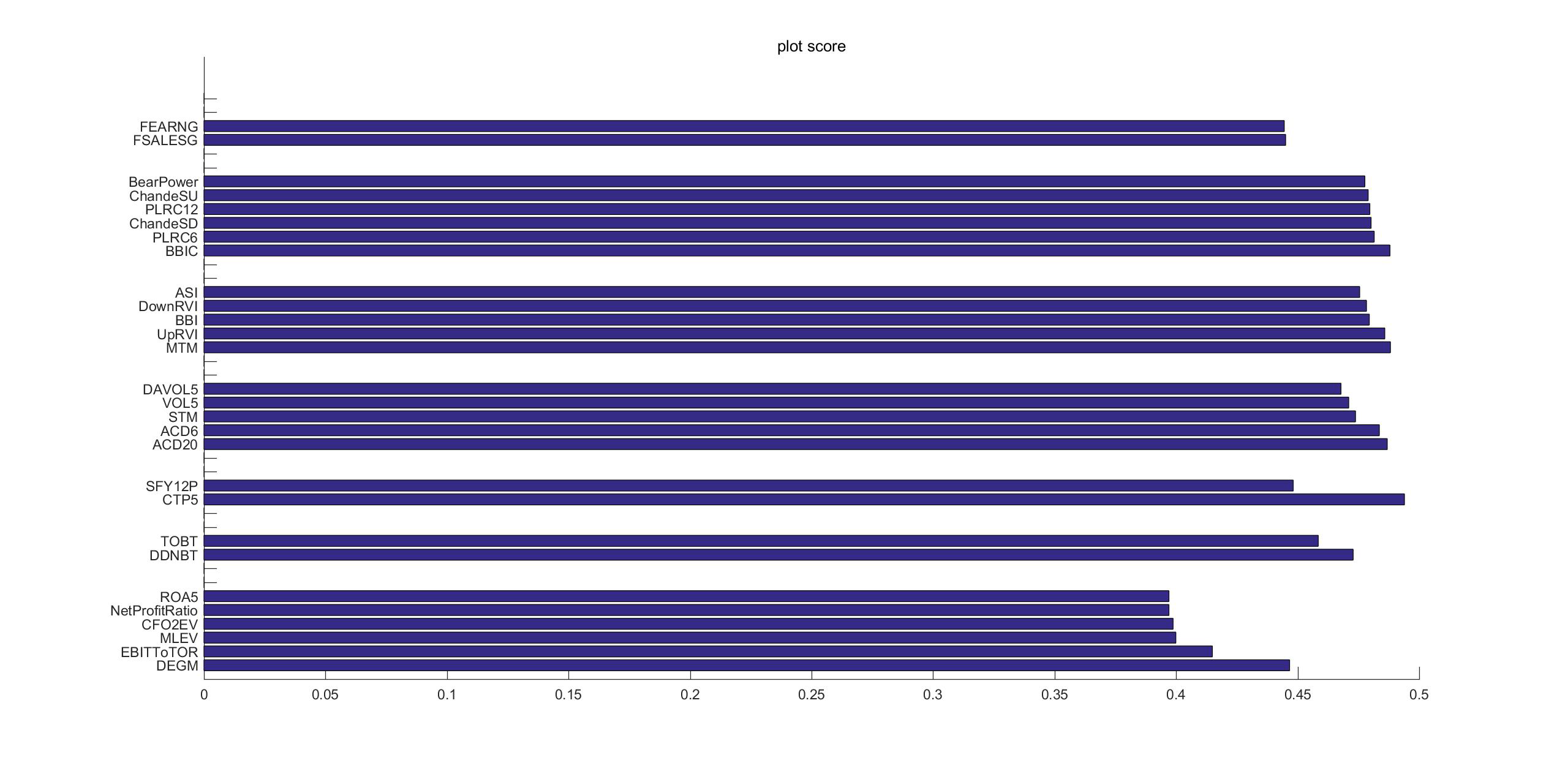}
	\caption{Score of Features} \label{fig 3}
\end{figure*}

\section{GMM-HMM}
\subsection{\underline{Introduction}}
Gaussian Mixture-Hidden Markov Model (GMM-HMM) has been widely used in automatic speech recognition (ASR) and has made much success.

Motivated by its application, our task is: given a observation sequence, find most likely hidden state sequence, and then find relation between hidden state and stock market trend by using Long-Short Term Memory.

Denote observation sequence $ O = O_1O_2...O_T$, where $O_t$ is the observation of time t. Hidden state sequence $ S = S_1S_2...S_T$, where $S_t$ is the hidden state of time t. The most likely hidden state is:

$$ S = argmax_S  P\{S|O\} $$

In this section, we will introduce core algorithms of GMM-HMM in predicting stock price and its practical operation

\subsection{\underline{Basic Hypothesis}}
Here are two basic assumptions when applying the model.
(1) The state at any time $t$ depends on its state at time $t-1$, and is independent of any other moment.
\begin{equation}
P \{S_t | S_{t-1},O_{t-1},...,S_1,O_1\} = P \{S_t | S_{t-1}\}
\end{equation}

(2) The observation of any time $t$ depends on its state at time $t$, and is independent of any others.
\begin{equation}
P \{O_t | S_T,O_T,...,S_1,O_1\} = P \{O_t |S_t\} 
\end{equation}

\subsection{\underline{GMM-HMM}}

GMM is a statistical model which can model $N$ sub population normally distributed. HMM is a statistical Markov model with hidden states. When the data is continuous, each hidden state is modeled as Gaussian distribution. The observation sequence is assumed to be generated by each hidden state according to a Gaussian mixture distribution [\hyperref[ref 1]{1}].

The hidden Markov model is determined by the initial probability distribution vector $\pi$, the transition probability distribution matrix A, and the observation probability distribution called emission matrix B. $\pi$ and A determine the sequence of states, and B determines the sequence of observations.

\begin{figure}[ht]
	\centering
	\includegraphics[scale=0.33]{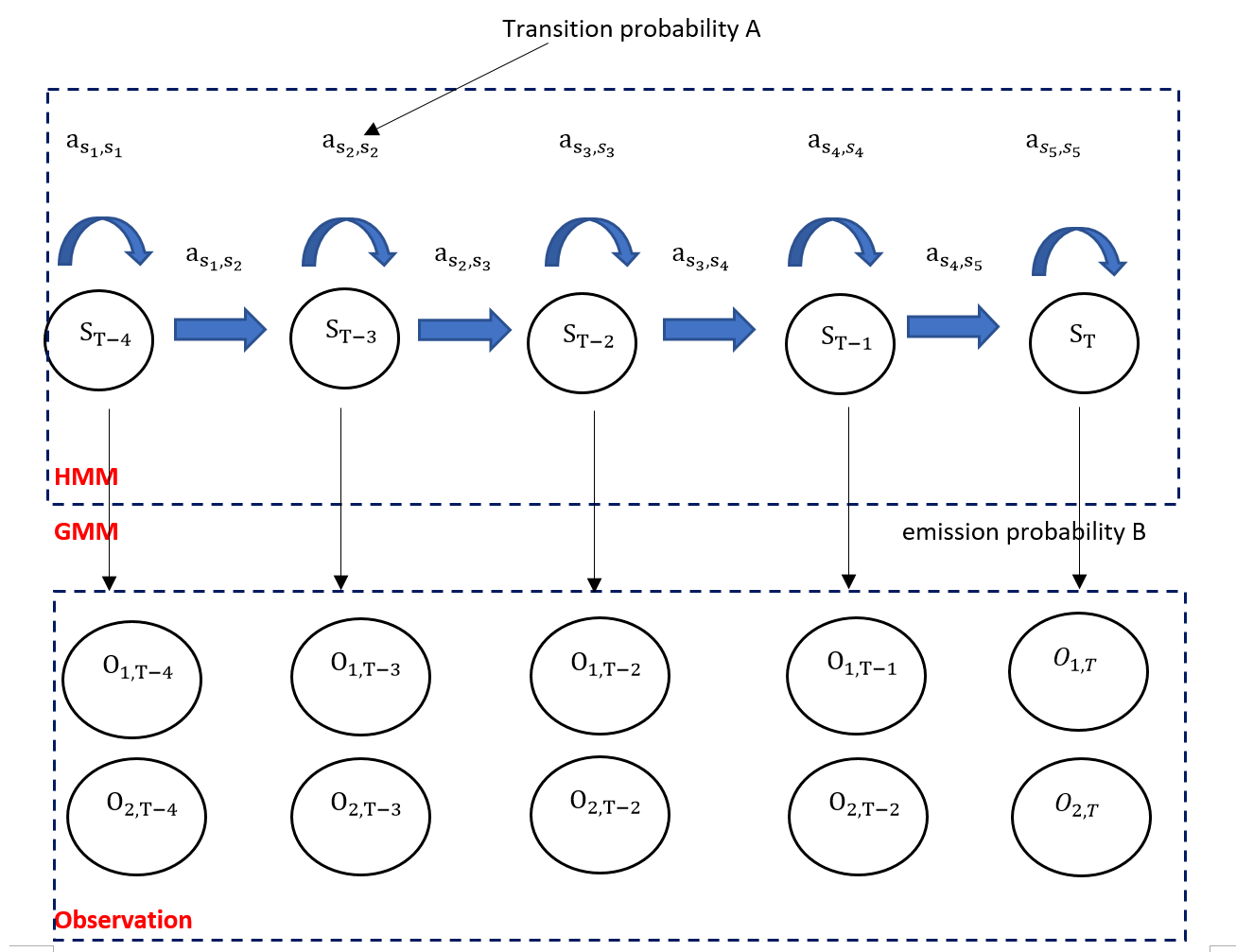}
	\caption{ GMM-HMM Structure} \label{fig 4}
\end{figure}

In GMM-HMM, the parameter B is a density function of observation probability, which can be approximated as a combination of products of mixed Gaussian distributions. By evaluating each Gaussian probability density function weight $w_{jk}$, mean and covariance matrix, the continuous probability density function can be expressed as:
\begin{equation}
b_j (V_t )= \sum_{k=1}^M w_{jk} b_{jk} (V_t ),j=1,...,N,0 \leq t \leq M-1 
\end{equation}
in which $w_{jk}$ represents the weight of each part. Fig. \ref{fig 4} describes the basic structure of GMM-HMM model.

\subsection{\underline{GMM-HMM Algorithm}}
Here we introduce the process of GMM-HMM algorithm.

\begin{enumerate}[1)]
\item{Determine the number of hidden states N}

\item{Estimate model parameters $\lambda = (A,B,\pi)$ by using Baum-Welh algorithm at a given observation sequence O. Record model parameters. Record model parameters $\lambda = (A,B,\pi)$ as gmm-hmm.}

\item{Estimate probability $P\{S_t = i \}, i = \{1,2,..,N\} ,t =\{0,1,..,T-1\}$ by using Viterbi algorithm in gmm-hmm.}

\end{enumerate}

\subsection{\underline{Results and Analysis of Experiment}}

In order to test the effect of our model, we select a
number of stocks for model training and determine the various
parameters of the model.

For better understanding, we choose one of the
stocks Fengyuan Pharmaceutical (000153.XSHE) which is
from 2007-01-04 to 2013-12-17 to do the visualization.

We set N=3, and use market data and predict the state at each moment. We use market information as a feature and set new model as \textit{gmm-mm1}.

\begin{figure*}[ht]
	\centering
	\includegraphics[scale=0.33]{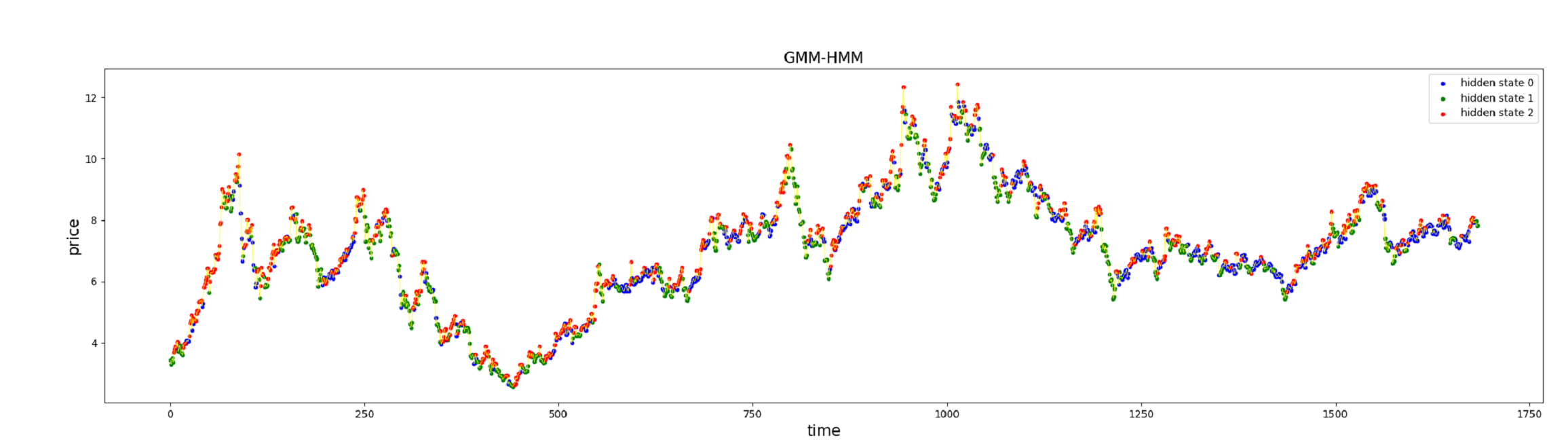}
	\caption{ XGB-HMM Traning} \label{fig 5}

	\centering
	\includegraphics[scale=0.3]{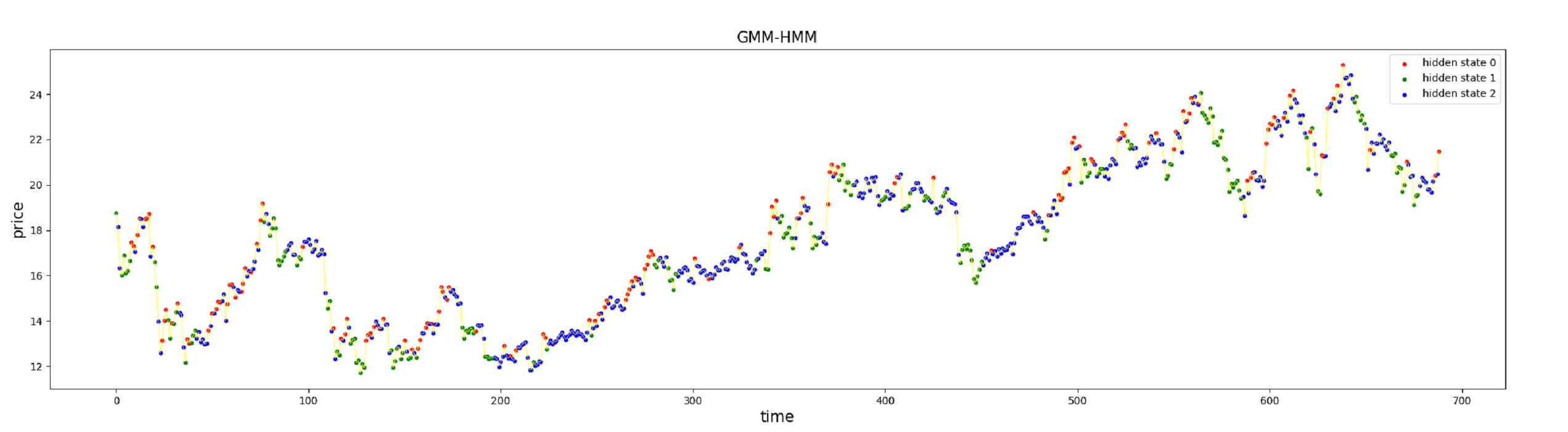}
	\caption{ XGB-HMM Test} \label{fig 6}

\end{figure*}

Then put the predicted state for time $t$ in \textit{gmm-mm1}
and its close price together to make visualization.
The results of the training set are shown in Fig \ref{fig 5} and \ref{fig 6}.

As can be seen from the figure, red represents the rising
state, green represents the falling state, blue represents the
oscillation, and intuitively \textit{gmm-mm1} works well on the
training set.

Use the data of the stock Red Sun (000525.XSHE) from
2014-12-01 to 2018-05-21 as a test set, use the predicted state
for time $t$ in \textit{gmm-hmm}1 and its close price together to
make visualization. From the Fig. \ref{fig 6} we can find that \textit{gmm-mm}1 works well on the test set.

Then, we trained seven multi-factor features that were
filtered to get the model $gmm-hmm_i$, $i = 2, 3,...8$, to
prepare for the latter LSTM model.

\section{XGB-HMM}
\subsection{\underline{Introduction}}
In the above GMM-HMM model, we assume that the
emission probability B is a Gaussian mixture model. Our team believes that in addition to the GMM distribution,
we can use the scalable end-to-end tree propulsion system
XGBoost to estimate the emission probability B.

In this section, we introduce a XGB-HMM model which
exploits XGBoost instead of Gaussian mixture model (GMM)
in estimating the emission probabiltiies $P\{O_t | S_t\}$. We will explain the training algorithms of XGBoost.

\subsection{Applying XGB into Emission Probability}
\subsubsection{XGB-HMM}

Machine learning and data-driven methods have become very important in many areas. Among the machine learning methods used in practice, gradient tree boosting is one technique that shines in many applications. 

The linear combination of trees can fit the training data well even if the relationship between the input and output in the data is complicated. Thus the tree model is a highly functional learning method. Boosting tree refers to the enhancement method which uses the addition model (that is, the linear combination of the basis functions) and the forward stagewise algorithm, and the decision tree as the basis function. It is an efficient and widely used machine learning method [\hyperref[ref 2]{2}].

Compared with the general GBDT algorithm, XGBoost has the following advantages [\hyperref[ref 6]{6}]:

\begin{itemize}
\item {Penalize the weight of the leaf node, which is equivalent to adding a regular term to prevent overfitting.}

\item {XGBoost's objective function optimization utilizes the second derivative of the loss function with respect to the function to be sought, while GBDT only uses the first-order information.}

\item {XGBoost supports column sampling, similar to random forests, sampling attributes when building each tree, training speed is fast, and the effect is good.}

\item { Similar to the learning rate, after learning a tree, it will reduce its weight, thereby reducing the role of the tree and improving the learning space.}

\item {The algorithm for constructing the tree includes an accurate algorithm and an approximate algorithm. The approximate algorithm performs bucketing on the weighted quantiles of each dimension. The specific algorithm utilizes the second derivative of the loss function with respect to the tree to be solved.}

\item {Added support for sparse data. When a certain feature of the data is missing, the data is divided into the default child nodes. }

\item {Parallel histogram algorithm. When splitting nodes, the data is stored in columns in the block and has been pre-sorted, so it can be calculated in parallel, which is, traversing the optimal split point for each attribute at the same time.}

\end{itemize}

\subsubsection{XGB-HMM Training Algorithm}

Referring to the GMM-HMM model training algorithm, our team has obtained the training algorithm of the XGB-HMM model as follows.

\begin{enumerate}[1)]
\item{Initialize:}
\begin{itemize}
\item{Train one GMM-HMM parameters $\lambda=(A,B,\pi)$, assuming the model is gmm-hmm.}

\item{Calculate $\alpha_t(i),\beta_t(i)$ by using forward-backward algrithm, and then calculate $\gamma_t(i)= P\{S_t = i |O,\lambda \}  $ and $ \gamma_t(i,j) = P\{x_t =q_i, x_{t+1}=q_j|O,\lambda\} $}
\vspace{0.5cm} for $t=1,2,..,T-1$ and $i = 0,1,..,N-1$

\begin{equation}
\alpha_t(i) = P\{O_0,O_1,...,O_t,x_t =q_i |\lambda \} 
\end{equation}

\begin{equation}
\beta_t(i) = P\{O_{t+1},O_{t+2},...,O_{T-1}|x_t =q_i ,\lambda \}
\end{equation}

\begin{equation}\gamma_t(i) = \frac{\alpha_t(i) \beta_t(i)}{P\{O|\lambda\}} 
\end{equation}

\begin{equation}
\gamma_t(i,j) = \frac{a_t(i)a_{ij}b_j(O_{t+1})\beta_{t+1}(j)}{P\{O|\lambda\}} 
\end{equation}
\end{itemize}
\item{Re-estimate:}
\begin{itemize}

\item{According
\begin{equation}
a_{ij} = \frac{\sum_{t=0}^{T-2} \gamma_t(i,j) }{\sum_{t=0}^{T-2}\gamma_t(i)} 
\end{equation}
Re-estimate transition probability A.
}

\item{Input O, supervised learning $ \gamma_t(i) = P\{S_t = i |O,\lambda \}$ training a XGB model, say $XGB_0$.}

\item{Use $XGB_0$ to estimate $ \hat{\gamma_t}(i)$, recorded as $ \gamma_t(i) _{new}$, and then use}

\begin{equation} b_j(k)  = \frac{\sum_{t\in\{0,1,..,T-1\} \& O_t=k}\gamma_t(j)_{new}}{\gamma_t(j)_{new}} 
\end{equation}
to re-estimate emission probability B.

\item{If $P\{O|\lambda\}$ increases,returen to 2, continuously re-estimate $\lambda$; otherwise finish training.}
\end{itemize}
\item{Finally, we get $\lambda=(A,B,\pi)$ and $XGB_0$}
\end{enumerate}

\vspace{0.5cm}

The training algorithm can be summarized as follows:

\begin{enumerate}[1)]
\item{Initialize, $\lambda=(A,B,\pi)$.}
\item{Compute $ \alpha_t(i), \beta_t(i),\gamma_t(i),\gamma_t(i,j)$.}
\item{Train XGB.}
\item{Re-estimate the model $\lambda=(A,B,\pi)$.}

\item{If $P\{O|\lambda\}$ increases, goto 2.}

\item{Of course, it might be desirable to stop if $P\{O|\lambda\}$ does not increase by at least some predetermined threshold and/or to set a maximum number of iterations.}

\end{enumerate}

\begin{figure}[ht]
	\centering
	\includegraphics[scale=0.35]{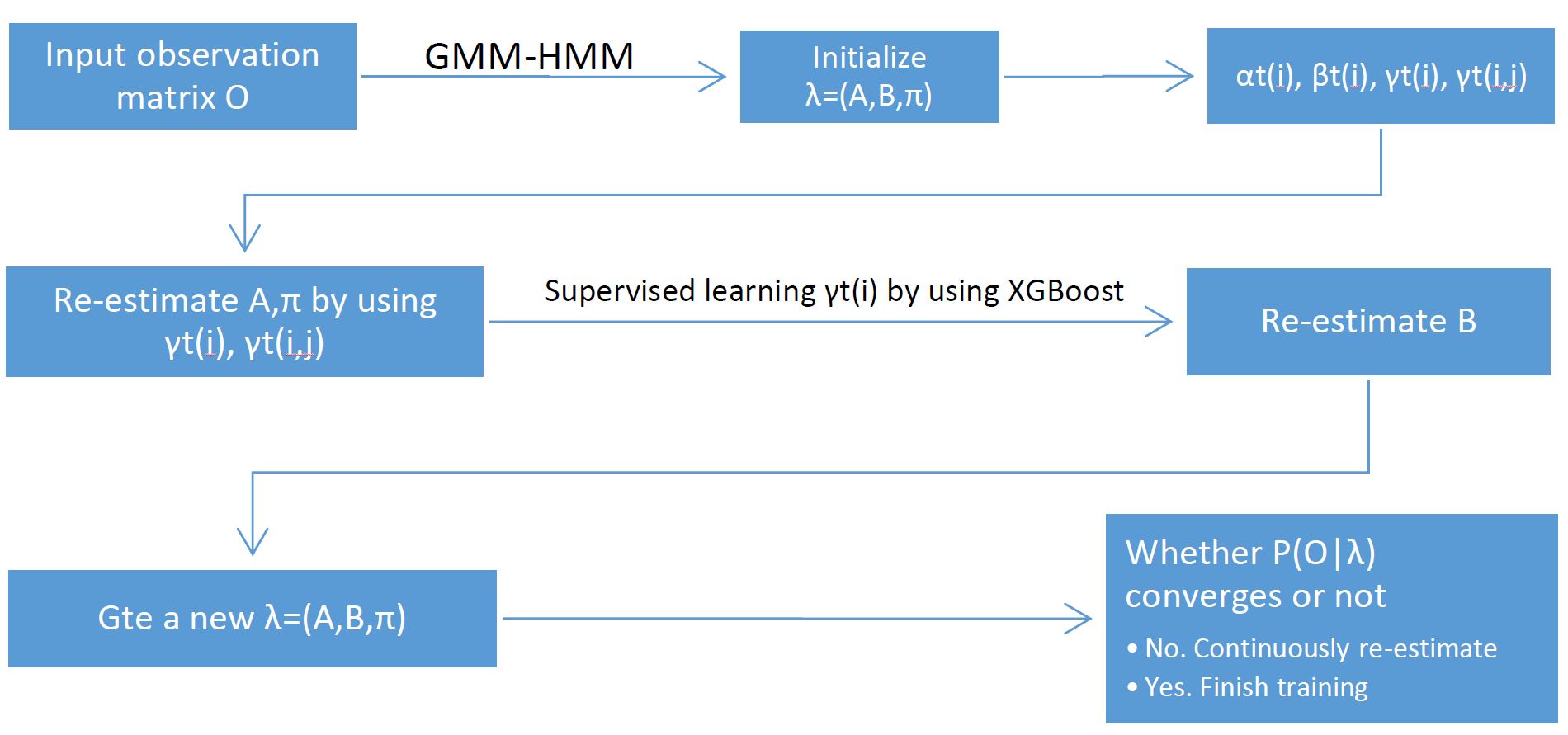}
	\caption{ XGB-HMM Training Algorithm} 
	\label{fig 7}
\end{figure}

XGB-HMM training algorithm in Fig. \ref{fig 7}.

\subsection{\underline{XGB-HMM Results Conclusion}}
In order to test the effect of our model, we select one of the stocks for model training and determine the various parameters of the model.

\begin{figure*}[ht]
	\centering
	\includegraphics[scale=0.32]{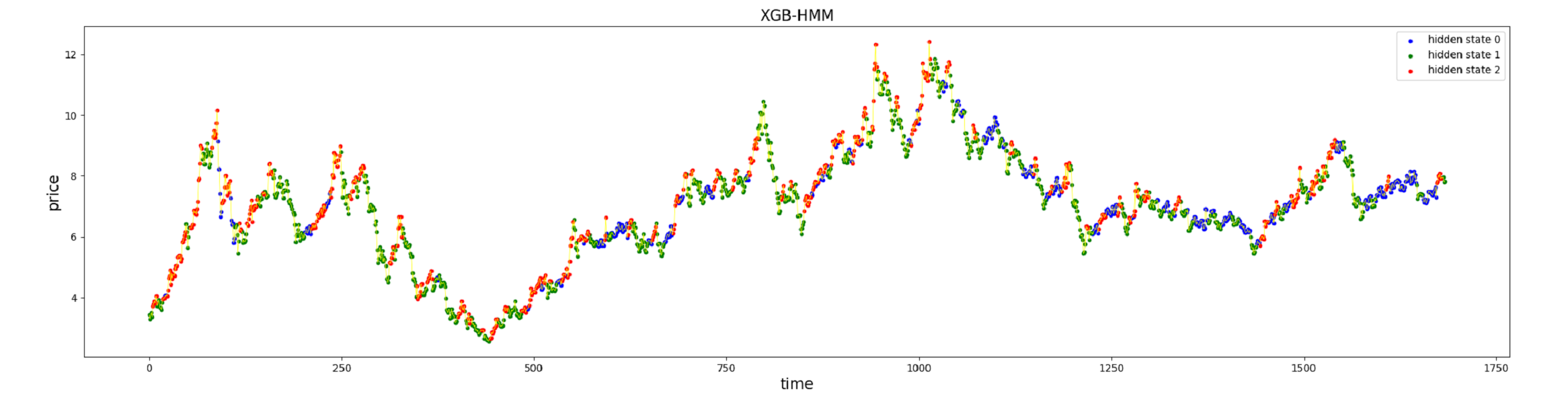}
	\caption{ XGB-HMM Traning} \label{fig 8}

	\centering
	\includegraphics[scale=0.3]{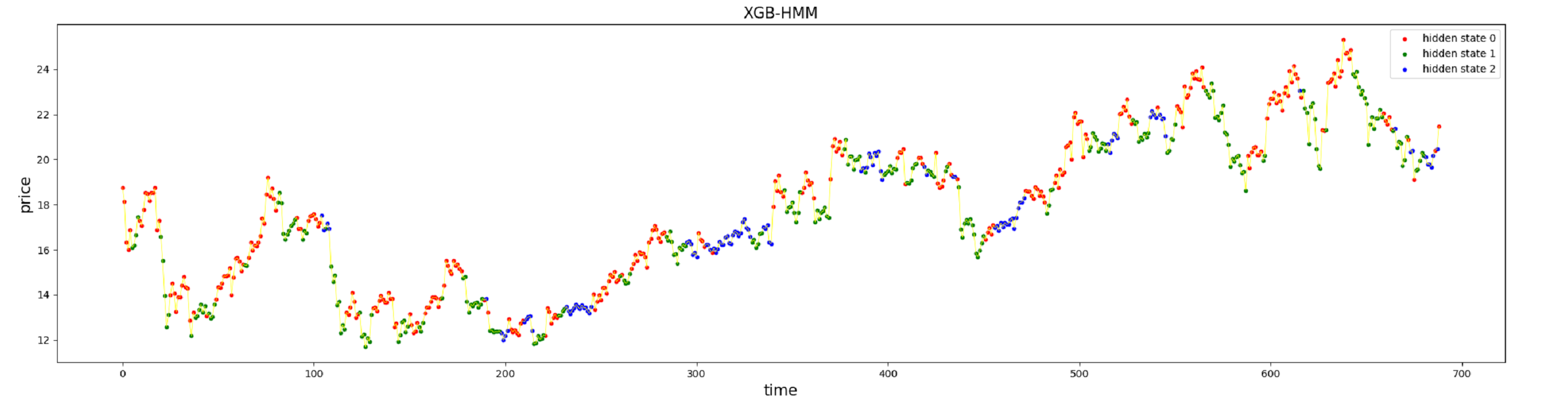}
	\caption{ XGB-HMM Test} \label{fig 9}

	\centering
	\includegraphics[scale=0.2]{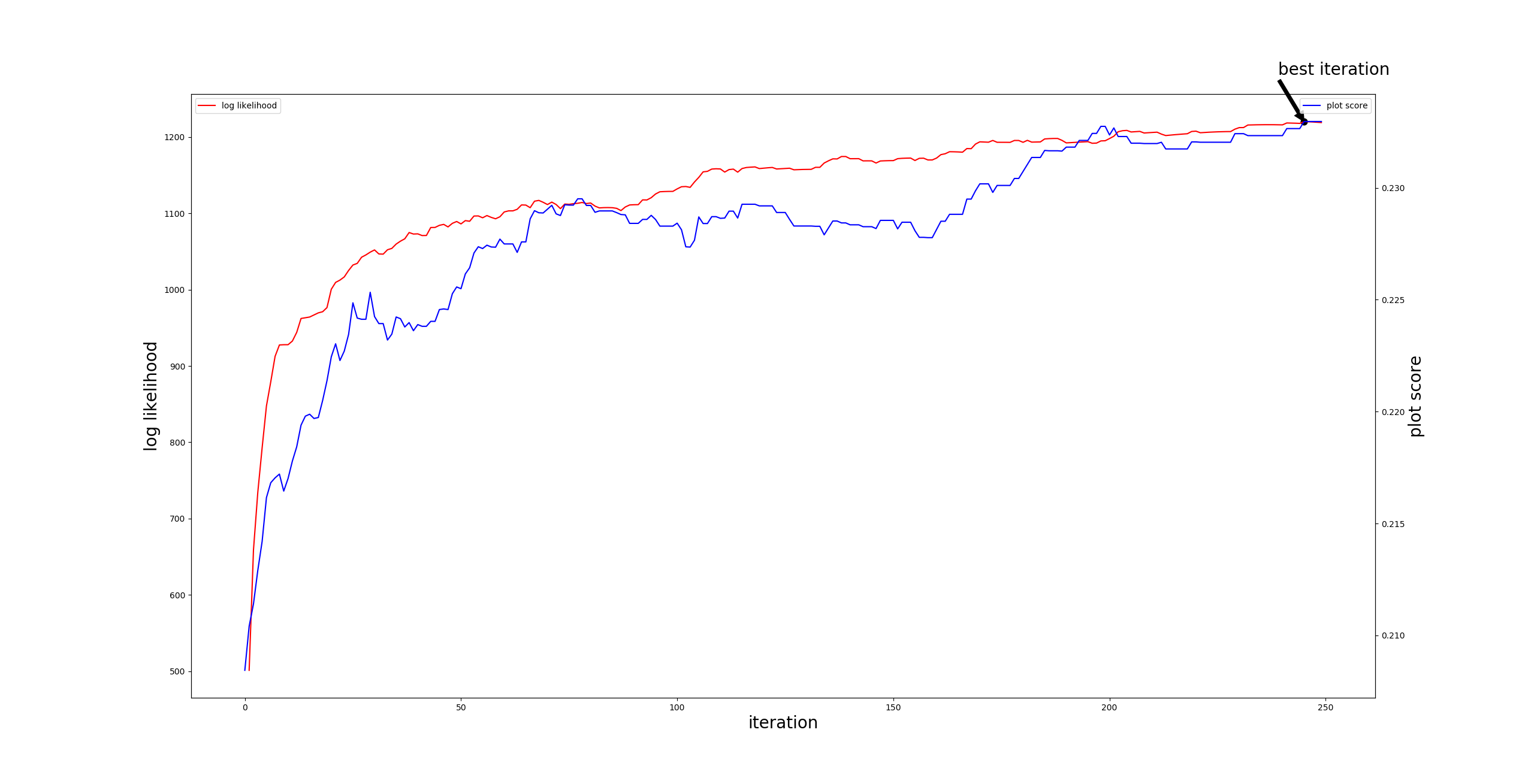}
	\caption{Iteration } \label{fig 10}
\end{figure*}

We choose Fengyuan Pharmaceutical (000153.XSHE), the training set is from 2007-01-04 to 2013-12-17. By setting N=3, and use market data as a feature, we predict the state at each moment. The results of the XGB-HMM training set are shown in Fig. \ref{fig 8}. The results of the XGB-HMM test set are shown in Fig. \ref{fig 9}. The iteration diagram of the XGB-HMM algorithm is shown in Fig. \ref{fig 10}.

As can be seen from the Fig. \ref{fig 8}, orange represents the rising state, green represents the falling state, blue represents the oscillation, and intuitively XGB-HMM works well on the training set, and the model is $xgb-hmm_0$.

Use the data of the stock Red Sun (000525.XSHE) from 2014-12-01 to 2018-05-21 as a test set, use model as $xgb-hmm_0$ see in Fig. \ref{fig 9}.

As can be seen from the Fig. \ref{fig 10}, as the number of iterations increases, log-likelihood increases as well. When iteration reach about 250, log-likelihood tends to be stable. At the same time, as the number of iterations increases, score function of features increases.

\subsection{\underline{Comparison of GMM-HMM and XGB-HMM}}

\begin{figure*}[ht] 
	\centering
	\includegraphics[scale=0.55]{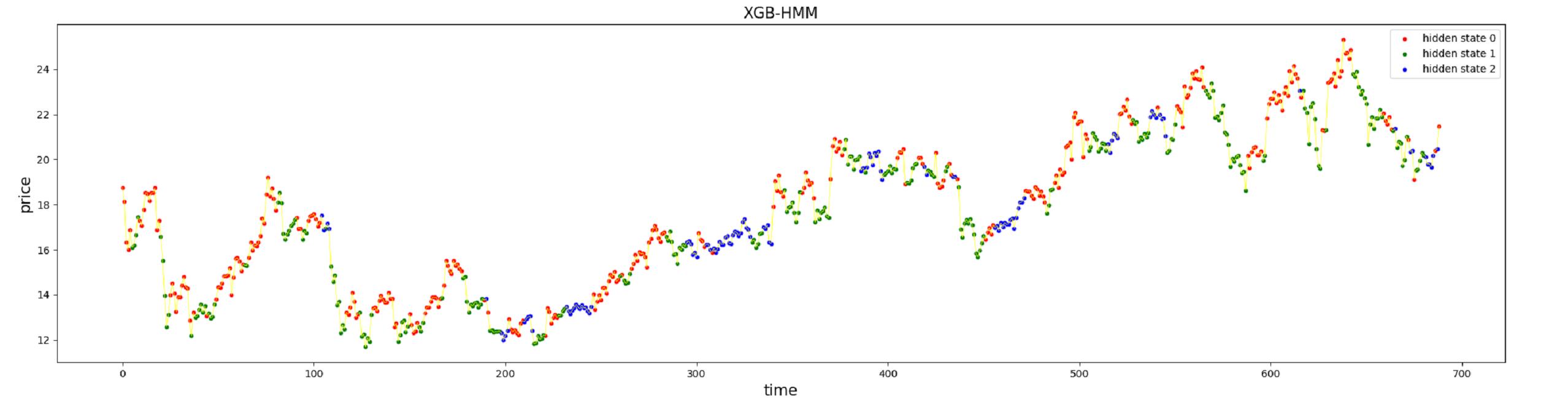}
	\caption{XGB-HMM Train} \label{fig 11}

	\centering
	\includegraphics[scale=0.62]{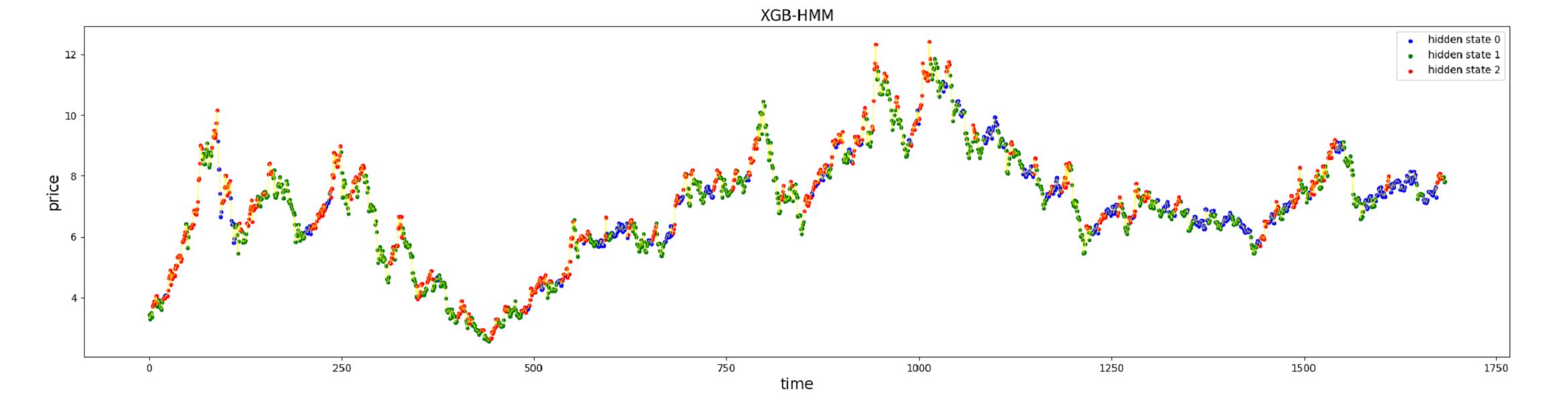}
	\caption{XGB-HMM Test} \label{fig 12}
\end{figure*}

Fig. \ref{fig 11} and \ref{fig 12} is the comparison of GMM-HMM and XGB-HMM results on test set and training set.

It can be seen that the result of using the XGB-HMM model is better, which makes the discrimination of the three hidden states higher, which means that XGB may have a better effect in fitting the observed data.

Moreover the results of XGB-HMM tends to be better than GMM-HMM.

\subsection{\underline{Strength and Weakness of Model and its Improvement}}

\subsubsection{Strength}

GMM cannot capture the relationship amoung different observation features, but XGB can.

In the training set, the accuracy of XGB training is 93\%. On the test set, the accuracy of XGB training is 87\%, and the fitting effect of XGB on the emission probability B is better than that of GMM.

Therefore, the XGB-HMM works better than the GMM-HMM in both the training set and the test set.

\subsubsection{Weakness}

In the GMM-HMM model and the XGB-HMM model, we use visualization to roughly observe the relationship between the state sequence S and the Y feature. It is considered that the red state represents the rise, the green state represents the decline, and the blue state represents the oscillation. 

However based on the training set of XGB-HMM and the test set results, we can see that the three states and the ups and downs of stock price are not relatively consistent. For example, when the stock price reaches a local maximum and starts to fall, the state is still red, and it is green after a few days.

Our team believes that the previous model did not take into account the probability of the state corresponding to each node. For example, when the state corresponding to a node is state 1, we think that this node must be state 1. But according to the XGB-HMM model, we can get the probability of each node corresponding to each state $P\{S_t = i \},i=1,2,3 $, and then get the probability matrix state\_proba.

\[ state\_proba= \begin{pmatrix} P\{S_1 = 1 \}......P\{S_{T-1} =1\}  \\  P\{S_1 = 2 \}......P\{S_{T-1} =2\} \\  P\{S_1 = 3 \}......P\{S_{T-1} =3\} \end{pmatrix} \]

Next we will use Long-Short Term Memory to find the relation between state\_proba and Y.

\vspace{0.5cm}

\section{GMM-HMM+LSTM \& XGB-HMM+LSTM}
\subsection{\underline{Introduction}}

The algorithm was first published by Sepp Hochreiter and Jurgen Schmidhuber on Neural Computation. The Long Short Term Memory model consists of a dedicated memory storage unit that controls the state of each memory storage unit through carefully designed forgetting gates, input gates and output gates. The control of each gate ensures the previous information can continue to propagate backwards without disappearing while the hidden layer is continuously superimposed with input sequence in the new time state [\hyperref[ref 5]{5}].

LSTM can be regarded as a mapping from X to Y, in which X is a n $ \times $ k matrix, Y is a vector with n columns, different X can map to one Y. 

\begin{figure}[H]
	\centering
	\includegraphics[scale=0.32]{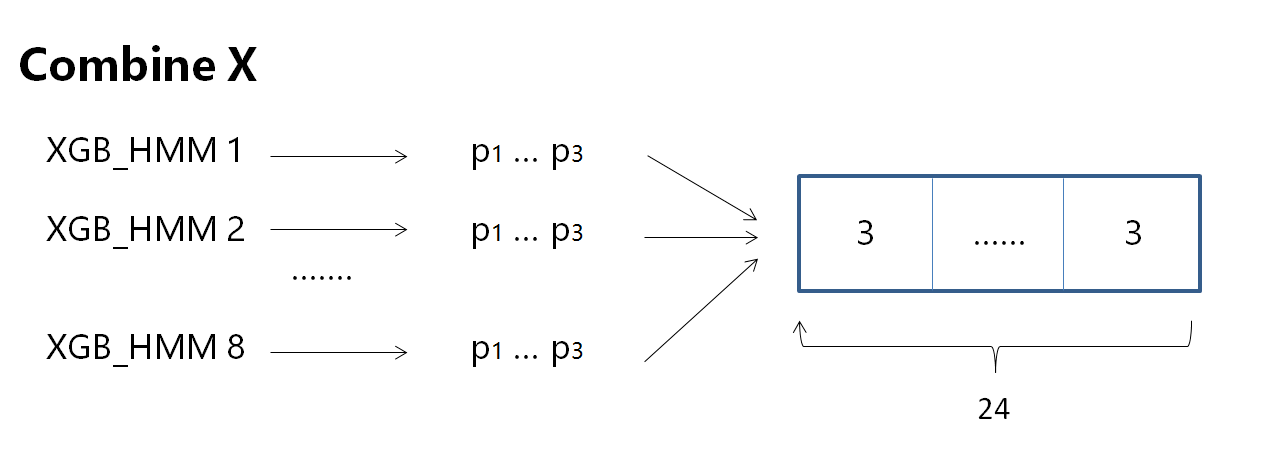}
	\caption{ LSTM} \label{fig 13}
\end{figure}
\vspace{0.5cm}

\subsection{\underline{Symbol}}

The symbols used in the figure have the following meanings:

\begin{table}[H]
	\centering
	\begin{tabular}{ll}
		\hline
		\textbf{Symbol}\\
		\hline
Type & Meaning \\
\hline
X & Information \\
\hline
+ & added information\\
\hline
$\sigma$ & Sigmoid layer \\
\hline
tanh & tanh layer \\
\hline
h(t-1) & the output of the previous LSTM unit \\
\hline
 c(t-1)& memory of the previous LSTM unit \\
\hline
 X(t)  & current input \\ 
\hline
c(t) & newly updated memory \\ 
\hline
h(t) & current output \\
		\hline\\
	\end{tabular}
	\caption{Symbol}\label{Table 4}
\end{table}

\begin{figure}[H]
	\centering
	\includegraphics[scale=0.28]{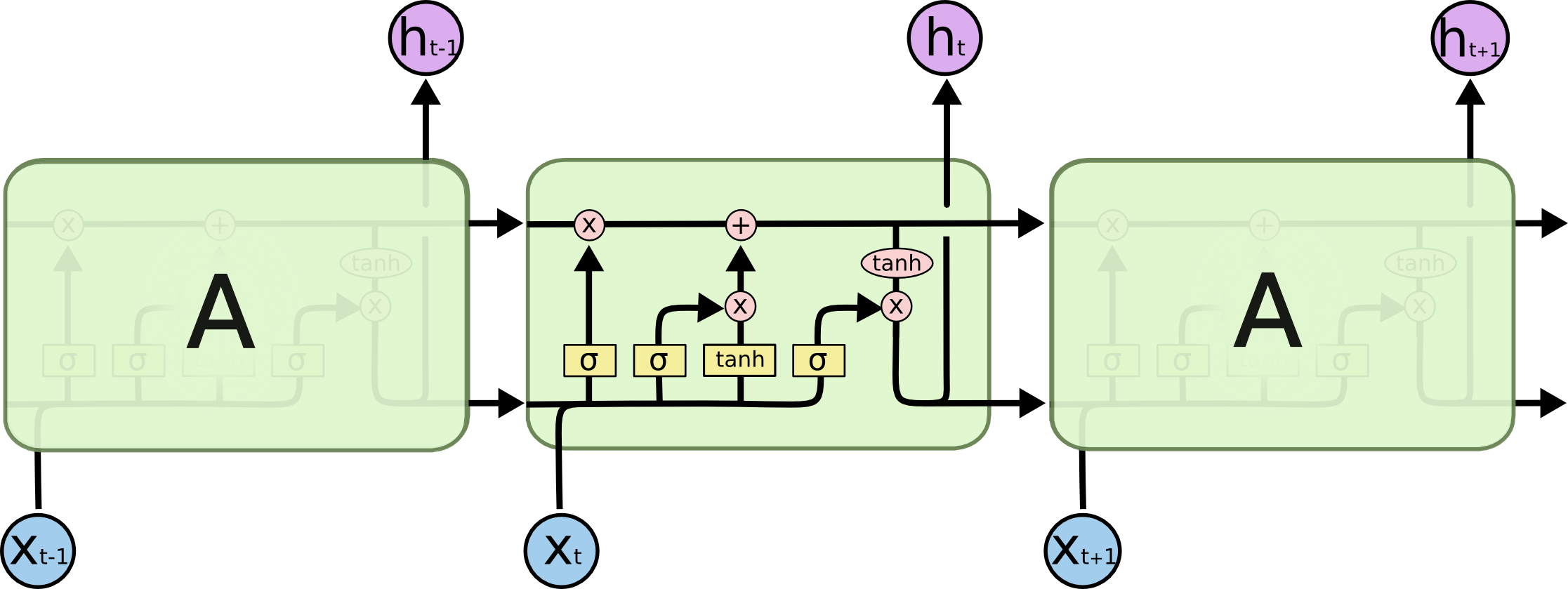}
	\caption{ LSTM Process} \label{fig 14}
\end{figure}

The long and short memory network structure is mainly composed of three parts:

\begin{enumerate}[1)]

\item{Forget gate layer.}
The first step in LSTM is to decide what information should be thrown away from the cell state. The decision is made by a sigmoid layer called ‘forget gate layer’. After linearly combining the output information $h_{t-1}$ of the previous layer and the current information $X_t$, the function value is compressed by the activation function to obtain a threshold value between 0 and 1. The closer the function value is to 1, the more information the memory retains. The closer to 0, the more information the memory loses [\hyperref [ref 3]{3}].

\begin{figure}[H]
	\centering
	\includegraphics[scale=0.3]{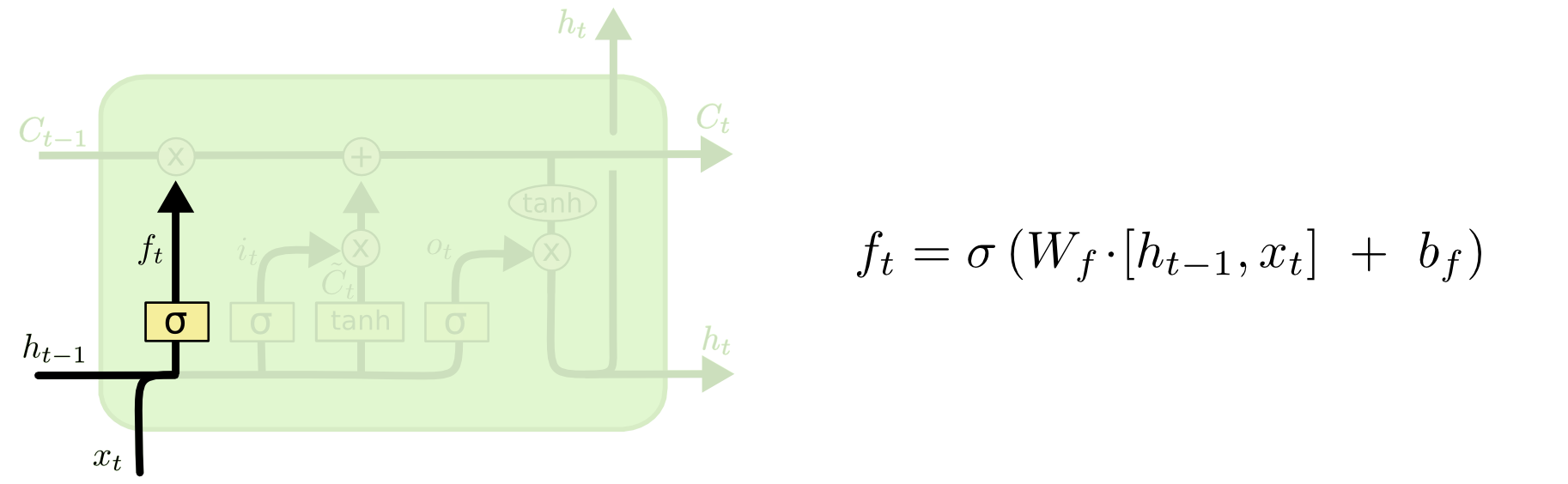}
	\caption{Forget Gate Layer[\hyperref [ref 3]{3}]} \label{fig 15}
\end{figure}

\item{Input gate layer.}
The input gate determines how much new information is added to the unit state. There are two steps to accomplish this: First, the sigmoid neural layer of the input gate determines which information needs to be updated; a tanh layer generates a vector that is used as an alternative to update the content $C_t$. Then, we combine the two parts and update the unit state.

\begin{figure}[H]
	\centering
	\includegraphics[scale=0.3]{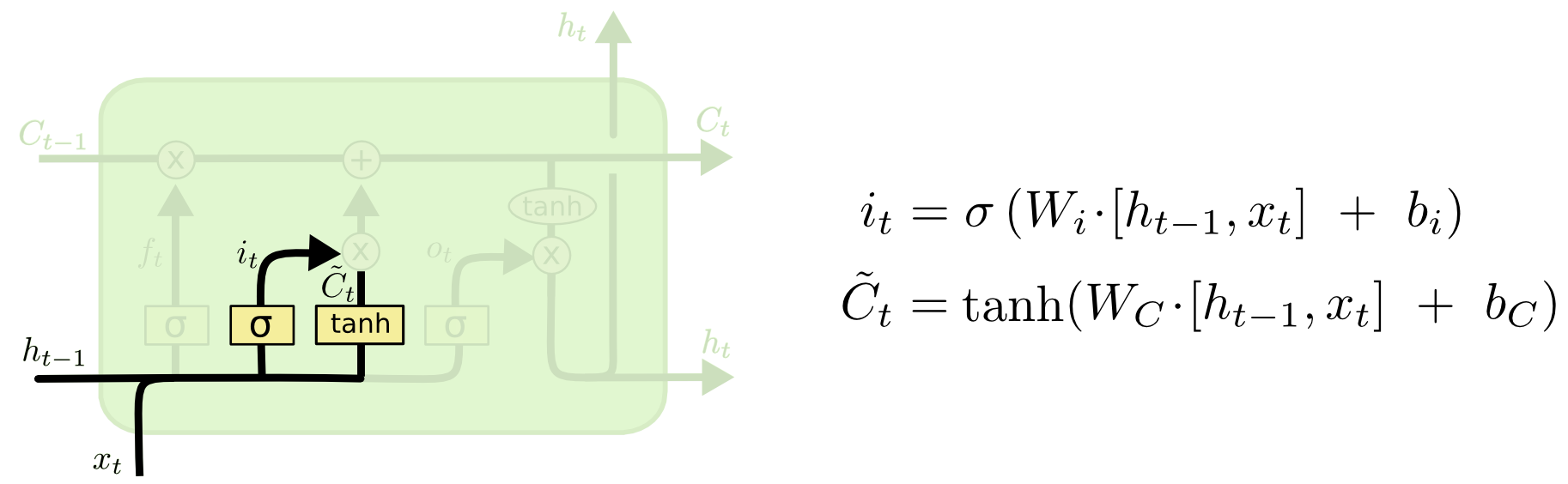}
	\caption{Input Gate Layer[\hyperref [ref 3]{3}]} \label{fig 16}
\end{figure}

\begin{figure}[H]
	\centering
	\includegraphics[scale=0.3]{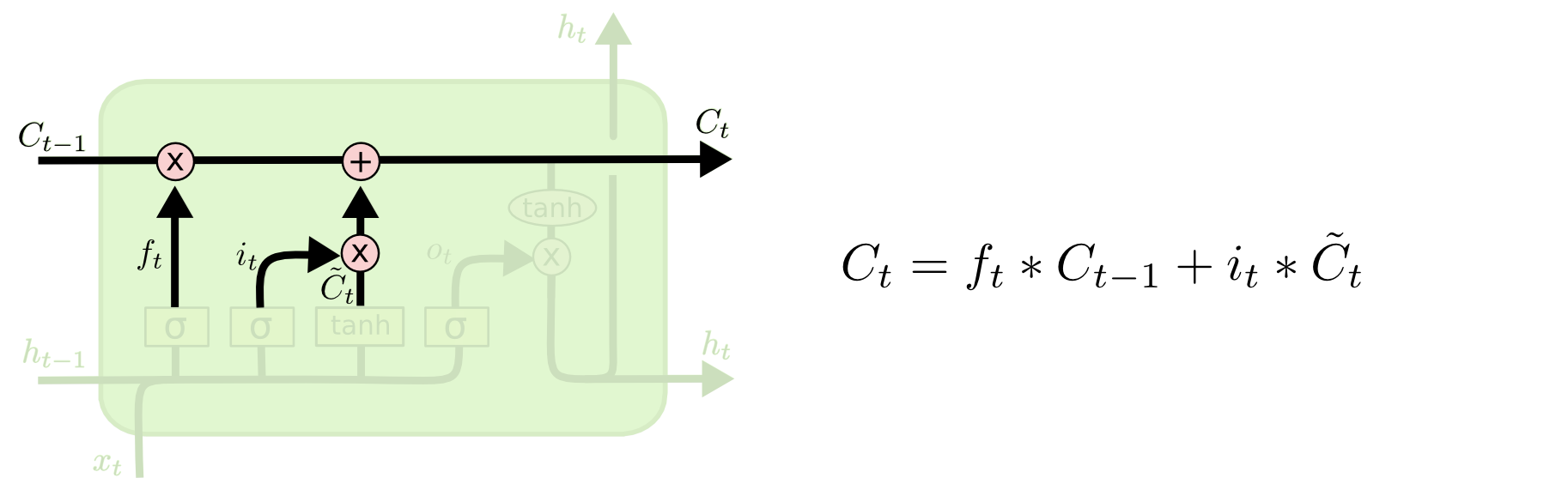}
	\caption{Input Gate and Candidate Gate[\hyperref [ref 3]{3}]} \label{fig 17}
\end{figure}

\item{Output gate layer.}
Finally, the output gate determines what value to output. This output is mainly dependent on the unit state $C_t$ and also requires a filtered version. First, the sigmoid neural layer determines which part of the information in $C_t$ will be output. Next, $C_t$ passes a tanh layer to assign values between -1 and 1, and then multiplies the output of the tanh layer by the weight calculated by the sigmoid layer as the result of the final output [\hyperref[ref 3]{3}].

\begin{figure}[H]
	\centering
	\includegraphics[scale=0.3]{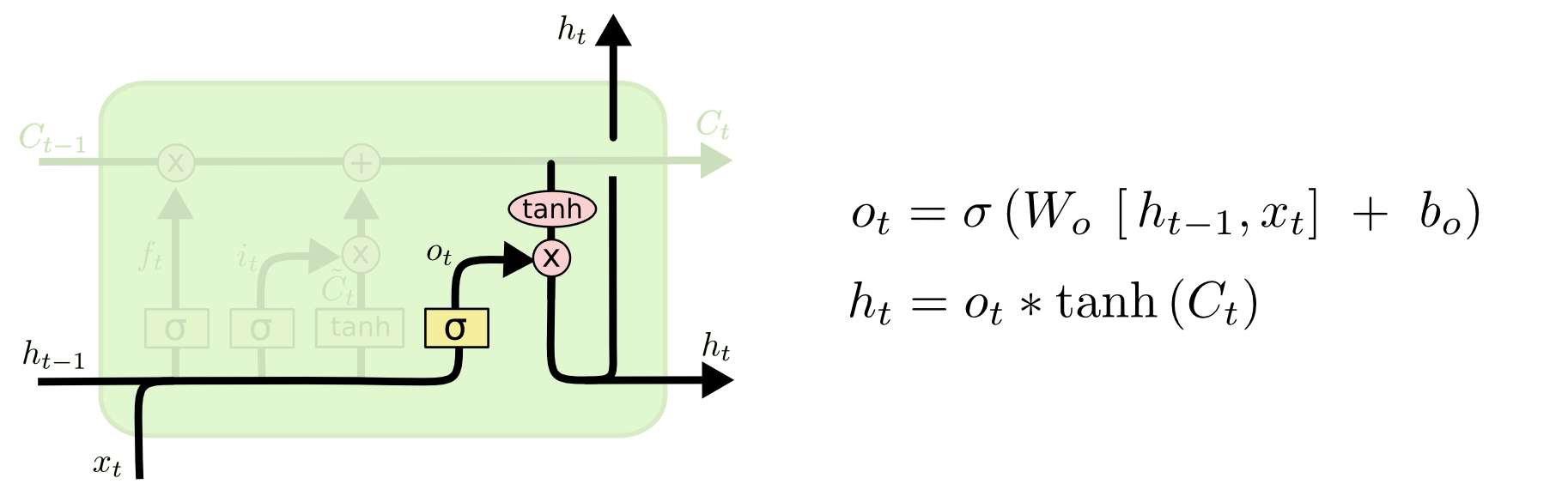}
	\caption{Output Gate Layer[\hyperref [ref 3]{3}]} \label{fig 18}
\end{figure}

\end{enumerate}
\subsection{\underline{Costruction of X in GMM-HMM+LSTM}}
On the  previous work, we have done $gmm-hmm_i,i=1,2,...,8$. For each $gmm-hmm_i,i=1,2,...,8$, we get one $state\_proba,i=1,2,...,8$, and then put them together to get X.

\subsection{\underline{Training algorithm}}
\begin{enumerate}[1)]
\item{Run GMM-HMM,and get $P\{S_t = i \}, i = \{1,2,..,N\},t =\{0,1,..,T-1\} $}

\item{Use probability $P\{S_t = i \}, i = \{1,2,..,N\},t =\{0,1,..,T-1\} $ to construct the training set X for LSTM}

 \[ X = \begin{pmatrix} P\{S_1 = 1 \}......P\{S_{T-1} =1\}  \\  P\{S_1 = 2 \}......P\{S_{T-1} =2\} \\  P\{S_1 = 3 \}......P\{S_{T-1} =3\} \end{pmatrix} \]

\item{Input X and Y into LSTM training}
\end{enumerate}

\subsection{\underline{Results and Analysis}}

The stock selected by our group is Fengyuan Pharmaceutical, and the data from 2007-01-04 to 2013-12-17 is selected as the training set. The stock Red Sun data from 2014-12-01 to 2018-05-21 is selected as a test set.

The GMM-HMM model is trained on 8 training factors on the training set, and 8 state\_proba matrices are generated to obtain X.

Start the training by entering this X into the LSTM model. The model completed in this training is $gmm-hmm+lstm_0$.

On the test set, run the $gmm-hmm+lstm_0$ model and output the result of the lstm model: the accuracy rate is 76.1612738\%.

\subsection{\underline{Construction of X in XGB-HMM+LSTM model}}
Above, our team has constructed the model $xgb-hmm_i, i=1, 2,...,8$. For each $xgb-hmm_i, i=1, 2,...,8$, you can get a $state\_proba,i=1,2,...,8$, and then merge them can get the matrix X.

\subsection{\underline{Training algorithm}}
\begin{enumerate}[1)]
\item{Run XGB-HMM, and get $P\{S_t = i \}, i = \{1,2,..,N\},t =\{0,1,..,T-1\} $}.

\item{Use probability $P\{S_t = i \}, i = \{1,2,..,N\},t =\{0,1,..,T-1\} $ to construct the training set X for LSTM.}

\[ X = \begin{pmatrix} P\{S_1 = 1 \}......P\{S_{T-1} =1\}  \\  P\{S_1 = 2 \}......P\{S_{T-1} =2\} \\  P\{S_1 = 3 \}......P\{S_{T-1} =3\} \end{pmatrix} \]

\item{Input X and Y into LSTM training.}

\end{enumerate}
\subsection{\underline{Results and Analysis}}

The stock selected by our group is Fengyuan Pharmaceutical, and the data from 2007-01-04 to 2013-12-17 is selected as the training set. The stock Red Sun data from 2014-12-01 to 2018-05-21 is selected as a test set.

The XGB-HMM model is trained on 8 training factors on the training set, and 8 state\_proba matrices are generated to obtain X.

Start the training by entering this X into the LSTM model. The model completed in this training is $xgb-hmm+lstm_0$.

On the test set, run the $xgb-hmm+lstm_0$ model and output the result of the lstm model: the accuracy rate is 80.6991611\%.

\subsection{\underline{Advantages and Disadvantages and Model Improvement}}

\subsubsection{Advantage}

LSTM is of time series.

In the final task of fitting state\_proba->label, we compared the effects of LSTM and XGB and found that LSTM is better than XGB.

\subsubsection{Disadvantage}

The accuracy of XGB on the training set is 93\%, and the accuracy on the test set is 87\%, which is better than the GMM fit, but there is room for improvement.

\subsubsection{Improve}

The processing of the data set and the construction of the features can be made more detailed.
Adjust the model parameters to make the final rendering of the model better.
\vspace{0.5cm}

\section*{Acknowledgement}
We would like to show gratitude to Guangzhou Shining Midas Investment Management Co., Ltd. for excellent support. They provided us with both in technological instructions for the research and valuable resources. Without these supports, it's hard for us to finish such a complicated task.

\vspace{0.5cm}



\end{document}